\documentclass{elsart3p}

\usepackage{natbib}

 \usepackage{graphics}
 \usepackage{epsfig}

\usepackage{amssymb}

\newcommand{\sSun}{ {\scriptscriptstyle{\rm \odot}} }

\begin{document}

\begin{frontmatter}


\title{Observational evidence of CMEs interacting in the inner heliosphere as inferred from MHD simulations}
\author[label1,label2]{N.~Lugaz \corauthref{cor1}},
\ead{nlugaz@umich.edu}
\author[label1]{W.~B.~Manchester~IV}, 
\author[label2]{I.~I.~Roussev},
\author[label1]{T.~I.~Gombosi}
\corauth[cor1]{Corresponding author. Institute for Astronomy, University of Hawaii, 2680 Woodlawn Dr., Honolulu, HI 96822, USA \newline Tel: +1 808 956 8534}
\address[label1]{Center for Space Environment Modeling, University of Michigan, 2455 Hayward St., Ann Arbor, MI 48109, USA}
\address[label2]{Institute for Astronomy, University of Hawaii, 2680 Woodlawn Dr., Honolulu, HI 96822, USA}

\journal{Journal of Atmospheric and Solar-Terrestrial Physics}
\volume{70}
\runauthor{N. Lugaz et al.}
\firstpage{598}
\FullCopyrightText{Accepted 29 August 2007 doi:10.1016/j.jastp.2007.08.033}

\begin{abstract}
The interaction of multiple Coronal Mass Ejections (CMEs) has been observed by LASCO coronagraphs and by near-Earth spacecraft, and it is thought to be an important cause of geo-effective storms, large Solar Energetic Particles events and intense Type II radio bursts. New and future missions such as STEREO, the LWS Sentinels, and the Solar Orbiter will provide additional observations of the interaction of multiple CMEs between the Sun and the Earth. 
We present the results of simulations of two and more CMEs interacting in the inner heliosphere performed with the Space Weather Modeling Framework (SWMF). Based on those simulations, we discuss the observational evidence of the interaction of multiple CMEs, both in situ and from coronagraphs. The clearest evidence of the interaction of the CMEs are the large temperature in the sheath, due to the shocks merging, and the brightness increase in coronagraphic images, associated with the interaction of the leading edges. The importance of having multiple satellites at different distances and angular positions from the Sun is also discussed.
\end{abstract}


\end{frontmatter}

\section{Introduction}\label{intro}
Coronal Mass Ejections (CMEs) are the most extreme events 
occurring in our solar system, and their frequency highly depends on the phase of the solar cycle: 
from 6 a day near solar maximum to 0.5-0.8 a day near solar minimum \citep[]{Gopalswamy:2004}.
The typical propagation time of a CME from the Sun to the Earth is 2-3 days. Therefore, near solar maximum, there is a high probability that multiple CMEs will interact on their way to Earth. Ejecta resulting from the interaction of multiple CMEs have been reported and studied by \citet{Burlaga:2002,Wang:2003, Berdichevsky:2003} and \citet{Farrugia:2004}, among others. 
Numerical investigations of multiple CMEs propagating and interacting, including the simpler case of the interaction between a CME and a forward shock-wave, have been pioneered by \citet{Vandas:1997} and recently reported by \citet{Odstrcil:2003b, Gonzalez:2004,Schmidt:2004, Lugaz:2005, Lugaz:2006} and \citet{Xiong:2006}. 

Based on near-Earth in situ measurements only, the interaction region between the magnetic sub-clouds of a multiple-magnetic cloud event is among the only evidence that multiple CMEs interacted between the Sun and the Earth. This region is characterized by a lower magnetic field strength, a higher temperature, resulting in a larger plasma $\beta$ (ratio of the thermal to the magnetic pressures) and it is associated with the reconnection between the two clouds \citep[]{Wang:2003}. As noted by \citet{Burlaga:2002}, the speed profile of complex ejecta, although irregular, often shows variations of less than 100 km~s$^{-1}$ between the different ejecta. Thus, it can be hard to distinguish between an isolated CME and interacting CMEs, simply based on the speed profile of the events observed near-Earth.

\begin{figure}[t*]
\begin{center}
\includegraphics*[width=7.5cm]{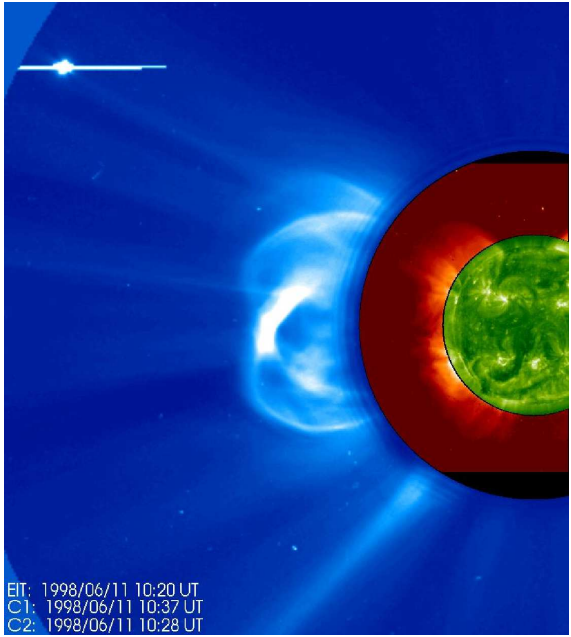}
\end{center}
\caption{Composite images (from SOHO/LASCO C1, C2 and EIT ) of two CMEs interacting in the solar corona on June 11, 1998. The two loops correspond to two sympathetic ejections.} 
\label{fig:LASCO1998}
\end{figure}

The interaction of two CMEs near the Sun can sometimes be observed by the LASCO coronagraphs \citep[e.g.][]{Gopalswamy:2001,Reiner:2003}. It can appear as CME  ``cannibalism" \citep[]{Gopalswamy:2001}, where the faster ejection ``swallows" the slower, preceding one. It can also appear as a brightness increase as the leading edge of the two CMEs interact, as is the case for the ejections from June 11, 1998 (see Figure \ref{fig:LASCO1998}).  However, often, the only indication that multiple CMEs interacted on their way to Earth is when multiple Earth-directed ejections are observed by LASCO and a single structure (multiple-magnetic cloud events or complex ejecta) is observed at Earth. 

It is the goal of this work to propose other evidence of CMEs interaction based on three-dimensional (3-D) magneto-hydrodynamic (MHD) simulations, relying both on existing ({\it Wind}, {\it ACE}, STEREO) and future missions (LWS Sentinels, Solar Orbiter).
We briefly summarize the simulations used for this study in Section \ref{model}. In Section \ref{1AU}, we discuss in situ synthetic measurements at 1 AU, followed, in Section \ref{LWS}, by a presentation of the possible in situ observations closer to the Sun by future missions. In Section \ref{STEREO}, we examine possible white-light observations of interacting CMEs by the STEREO Heliospheric Imagers. In Section \ref{conclusion}, we conclude and discuss other possible observational evidence of interacting CMEs not included in the present work. 

\section{Simulations of the Propagation and Interaction of Multiple CMEs Between the Sun and the Earth}\label{model}

The two simulations on which this study is based have been published in \citet{Lugaz:2005} and \citet{Lugaz:2006}. Both simulations are performed with a 3-D MHD code (BATS-R-US). In \citet{Lugaz:2005} (therafter Simulation 1), two identical out-of-equilibrium Gibson-Low magnetic flux ropes \citep[]{Gibson:1998} are added 10 hours after each other onto the solar surface into a solar wind characteristic of solar minimum \citep[see also][for a description of the models]{Manchester:2004b}. The interaction of those two ejections results in the passage of a multiple-magnetic cloud event at Earth. The two magnetic sub-clouds are preceded by a single shock wave, the result of the merging of the two shock waves driven by the ejections. In \citet{Lugaz:2006} (thereafter Simulation 2), we investigate three homologous eruptions from NOAA active region 9236 in November 24, 2000. The three ejections were separated by 10 and 6.5 hours respectively and of equivalent velocities (1000-1200~km~s$^{-1}$). We use the solar wind model developed by \citet{Roussev:2003} incorporating MDI magnetogram data and, which reproduces observations by {\it Wind} for the pre-eruption solar wind. We use out-of-equilibrium semi-circular magnetic flux ropes to initiate the three eruptions and are able to reproduce most of the LASCO and {\it Wind} observations. Simulation 2 was performed with the Space Weather Modeling Framework \citep[for a description of the SWMF, see][]{Toth:2005}.

\section{Synthetic Observations at 1 AU}\label{1AU}

\begin{figure*}[t*]
\begin{center}
\includegraphics*[width=17cm, height= 17cm]{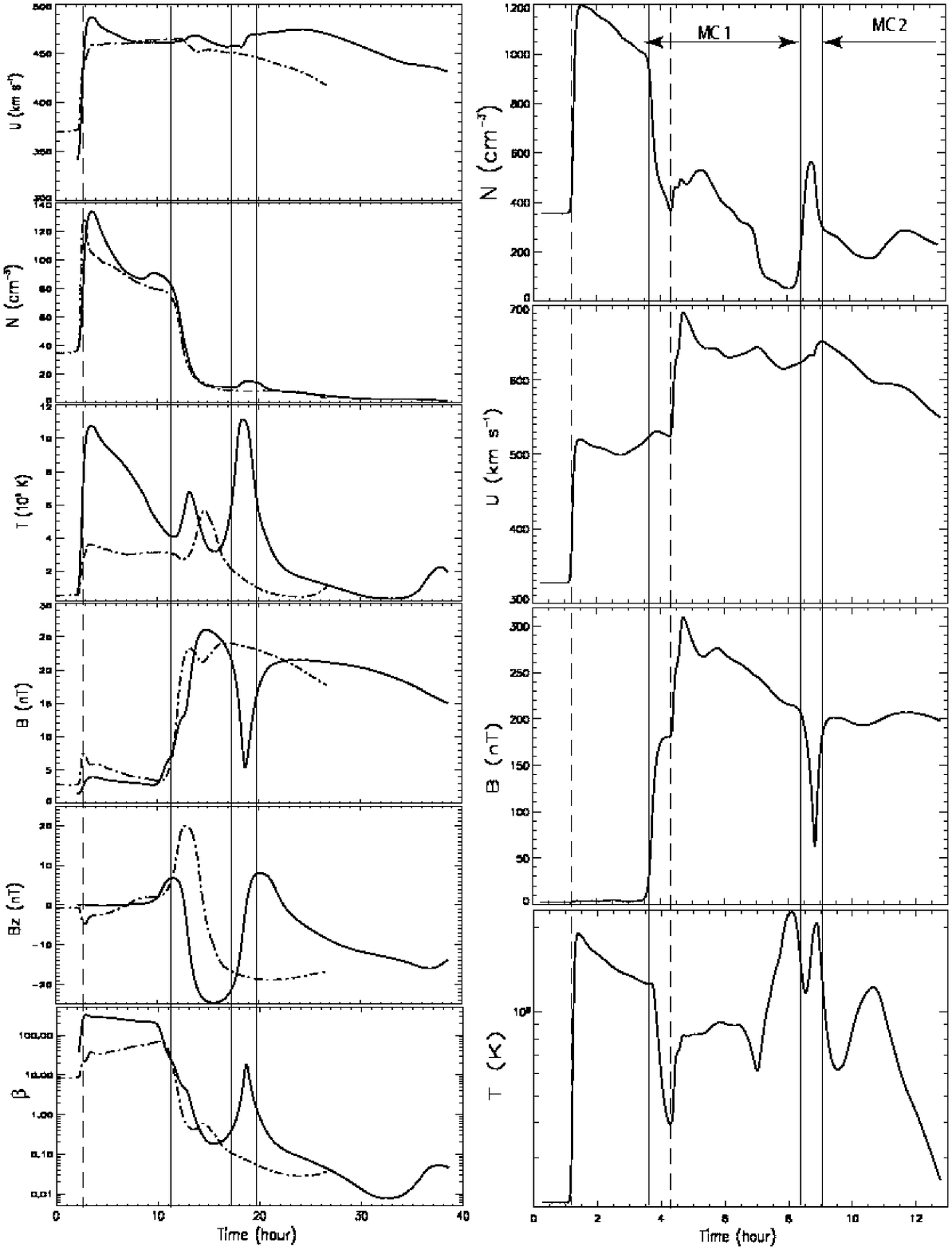}
\end{center}
\caption{{\it Left panel}: {\it Solid line}: Plasma properties near 1 AU for the two interacting CMEs.
{\it Dash-dotted line}: Plasma properties near 1 AU for the simulation by \citet{Manchester:2004b} of an identical isolated CME. The results are shifted in time in order to align the arrival of the shock fronts. \newline
{\it Right panel}: Plasma properties near 0.33 AU for the two interacting CMEs. In both panels, the solid vertical lines show approximately the boundary of the two magnetic clouds, and the dash vertical line(s) show the position of the forward shock(s).}
\label{fig:sat}
\end{figure*}

Here, we compare the results at 1 AU of Simulation 1 ({\it solid line}, thereafter referred as the interacting case) to the results of an identical but isolated CME ({\it dash-dotted line} from \citet{Manchester:2004b}, thereafter referred as the isolated case), as seen in the left panel of Figure \ref{fig:sat}. The goal is to find the differences at Earth between the simple superposition of two magnetic clouds and the result of the interaction of those two clouds.

In the interacting case, the single shock reaches 1 AU at time t = 66 hours, 6 hours before the shock in the isolated case. 
In order to directly compare the plasma parameters at 1AU, the results shown on the left panel of Figure \ref{fig:sat} are shifted in time, so that the arrival time of the shocks coincide. 

In the interacting case, after increasing by a factor of 3.6 across the shock, the
density decreases at first in the sheath, but then increases again
before the contact discontinuity at the interface with the first magnetic cloud.
This increase in density in the rear part of the sheath
is specific to the interacting case. This is because the rear part of the sheath has been shocked twice before the shocks' merging, whereas the front part is formed after their merging. Thus, the rear part of the sheath has been compressed by a factor greater than $(\gamma+1)/(\gamma -1)= 4$. However, the evolution of the density at the shock and in the sheath do not provide clear evidence of the  merging of two shocks.
Specifically in the interacting case, the maximum temperature is
reached just behind the shock and not inside the magnetic cloud. This
temperature jump at the shock is a factor of 2.5 larger than the
one found for the isolated case. This larger temperature is associated with the process of shocks' merging. Since the density and velocity jumps are found to be similar between
isolated and interacting cases, the temperature is the only clear indicator that the
shock observed at Earth is the result of the merging of two shocks. Also, the larger density in the sheath is expected to have important consequences for the geo-effectiveness of the event \citep[]{Farrugia:2006}.

One indication that the first magnetic cloud has been compressed and heated by a shock is that this cloud is hotter than the second one with a maximum temperature about 25$\%$ larger than that of the second cloud (and than that of the isolated cloud). However, this difference in temperature is smaller than the variation in temperature from one isolated magnetic cloud to the other. 
At Earth, the global structure composed by the two magnetic clouds
is about 50 $R_\sSun$ wide and the second magnetic cloud accounts for about $80 \%$
of this width. This structure is fairly different from the simple superposition of two identical
magnetic clouds. Also, the maximum magnetic field strength in the first cloud for the interacting case is only slightly larger (26 nT instead of 25 nT) than that of the isolated cloud. 
However, the maximum southward magnetic field in the first cloud reaches a maximum value of -24.7 nT (-20 nT in the isolated case). The maximum southward magnetic field of the second cloud is very similar to the one from the undisturbed case. One can expect the
geo-effectiveness of the interacting event to be larger, especially because the maximum southward $B_z$ is larger in the first cloud compared to the isolated case. 

The reconnection region between the two magnetic clouds is associated with values of $\beta$ above unity, corresponding to the low magnetic field and high plasma temperature. This region is associated with the reconnection of the two clouds, which results in magnetic energy being transformed into thermal energy. Such high-$\beta$ interaction regions are described by \citet{Wang:2003} for three different multiple-MC events in March-April 2001.

\section{Synthetic Observations Closer to the Sun}\label{LWS}

There has been no systematic in-situ observations of solar transients between the Sun and the L1 point since the days of Helios and IMP (1980s).

\begin{figure*}[t*]
\begin{minipage}[]{1.0\linewidth}
\begin{center}
{\includegraphics*[width=18cm]{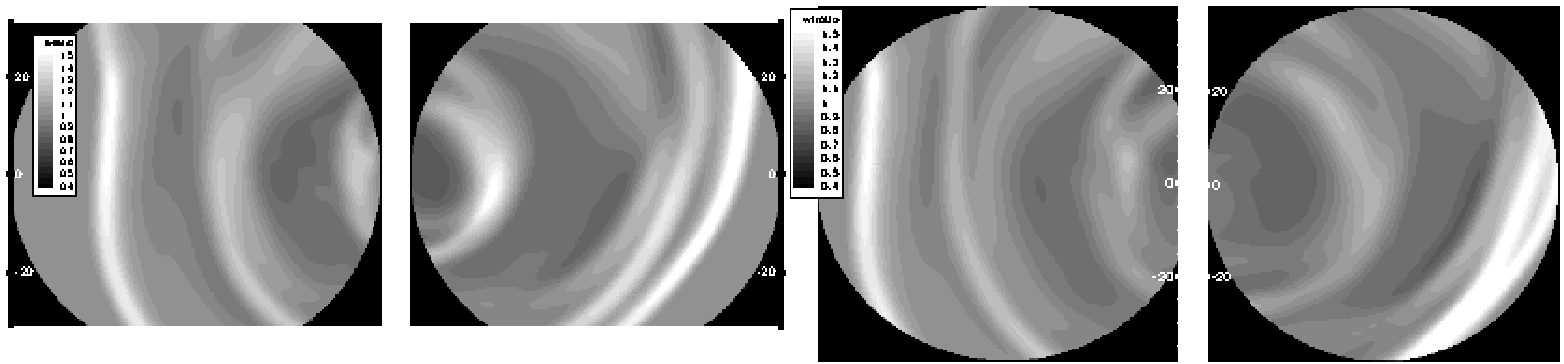}}\\
{\includegraphics*[width=6.cm]{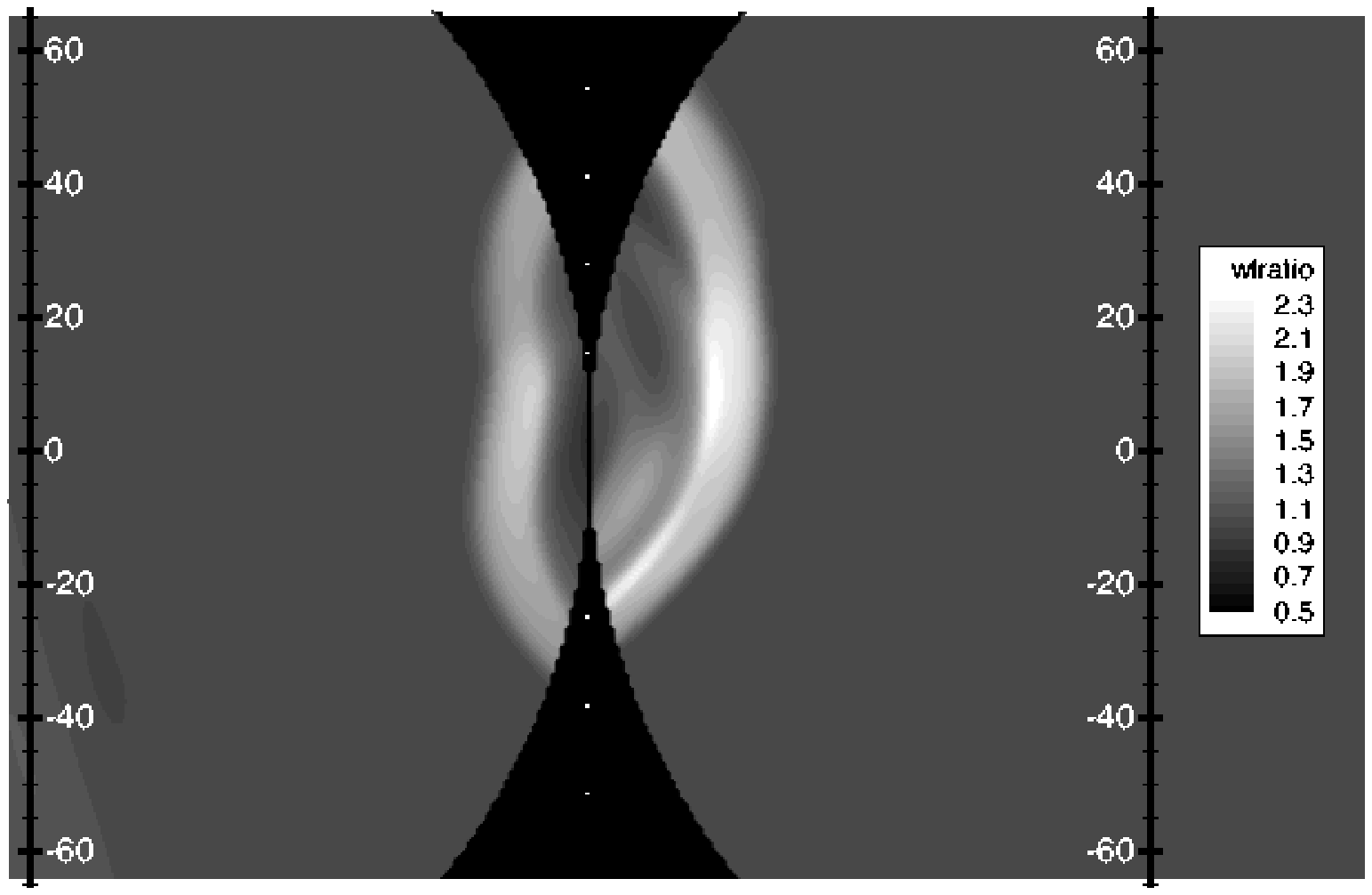}}
{\includegraphics*[width=5.cm]{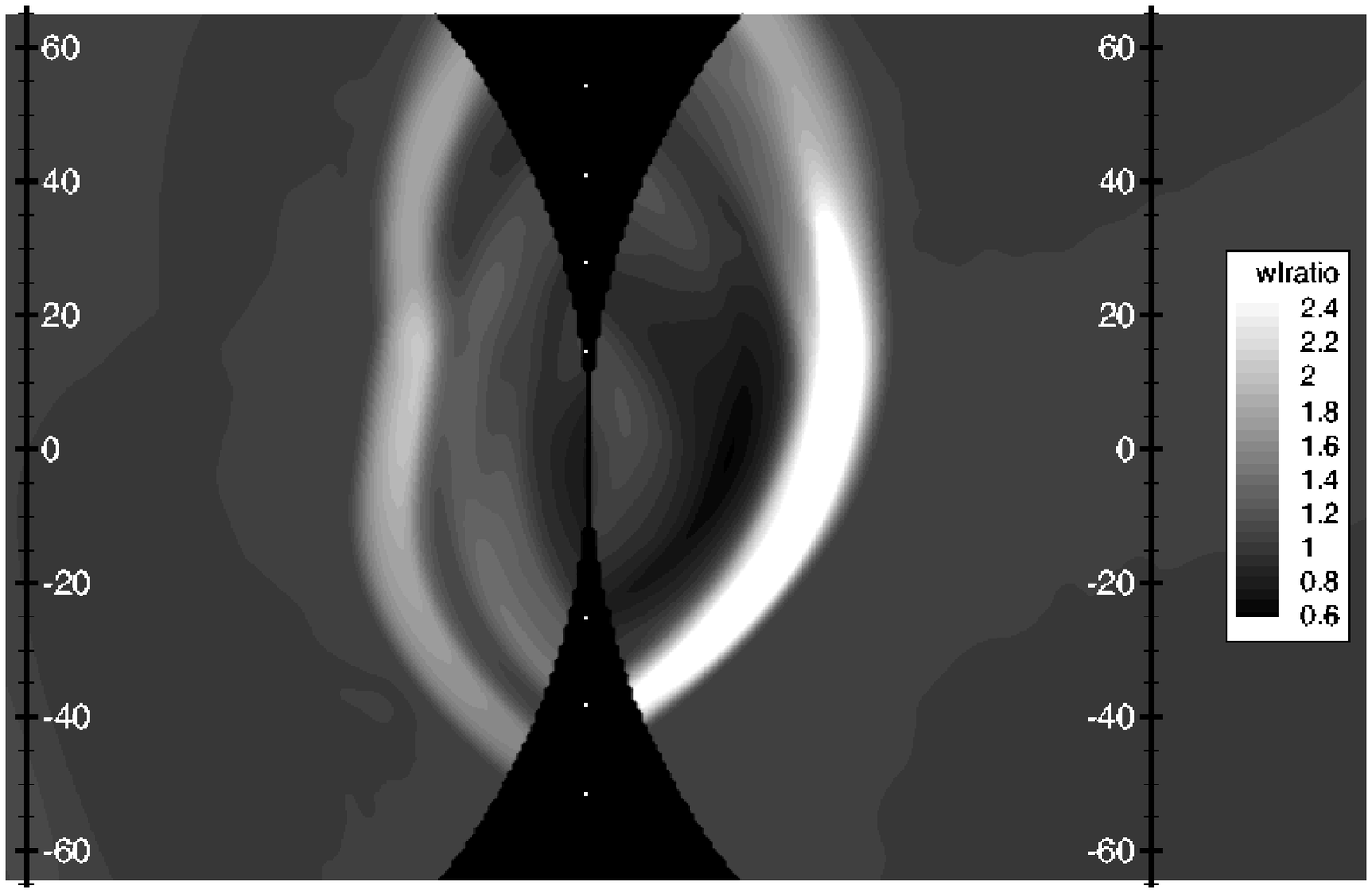}}
{\includegraphics*[width=5.cm]{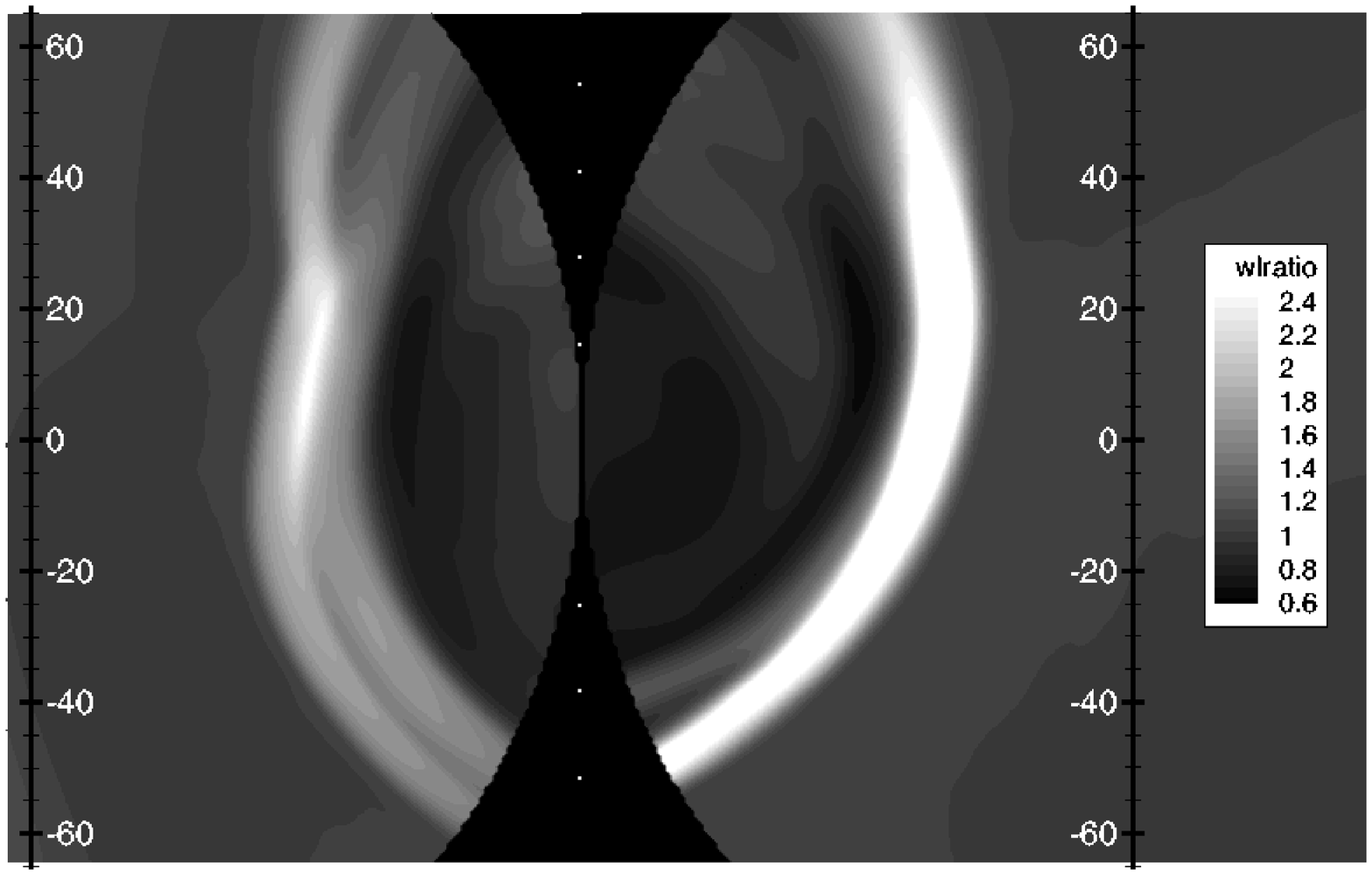}}
\end{center}
\end{minipage}\hfill
\caption{Synthetic STEREO Heliographic Imagers observations of the interaction of 3 CMEs. Running ratio of consecutive images. {\it Top}: Heliospheric Imagers 1, 22 and 25 hours after the initiation of the first CME (1-hour cadence). {\it Bottom}: Heliospheric Imagers 2, 22, 28 and 34 hours after the initiation of the first CME (3-hour cadence).}
\label{fig:HI1}
\end{figure*}

The Living With a Star (LWS) Sentinels is a project of four spacecraft to be launched in the second half of the 2010s into elliptical orbits between 0.3 and 0.8 AU. Solar Orbiter is planed to be launched by the mid-2010s into an elliptical orbit around the Sun reaching distance as close as 45~$R_\sSun$ from the Sun. This will provide unique opportunities to study shocks propagation and interaction with preceding ejecta. {\it ACE}, {\it Wind} and the two STEREO spacecraft will provide complementary in-situ observations near Earth's orbit.
Observations of shocks propagating into magnetic clouds are more likely to be made closer to the Sun than Earth's orbit. 
Near Earth, it is more likely to observe a shock having exited the magnetic cloud it overtook, as is the case for the March 20, 2003 event \citep[]{Berdichevsky:2005}. 
An example of observations closer to the Sun is the complex stream observed by Helios 2 in April, 2-6, 1979 as reported by \citet{Farrugia:2004}, where two shocks appear to propagate inside preceding ejecta. 

As an example of in-situ measurements close to the Sun, we consider the plasma properties of the two ejections from Simulation 1 as they would have been observed by a satellite along the Sun-Earth line at a distance of 0.33 AU from the Sun (see right panel of Figure \ref{fig:sat}). The results at 1 AU are shown with the solid line on the left panel of Figure \ref{fig:sat}. At 0.33 AU, the trailing shock 
is in the front part of the leading cloud
. Across the shock, the compression in density is about 1.32, the jump in the magnetic field strength is about 1.7, reflecting the modifications of the trailing shock properties by the extreme upstream conditions encountered \citep[see][section 4, for more details]{Lugaz:2005}. This is a case of shock propagating inside a magnetic cloud, also studied, for example, by \citet{Xiong:2006}.

The speed profile in the ejecta is not as uniform as observed at Earth, because the trailing shock is still propagating inside the first cloud and is still accelerating it. The temperature profile is also significantly different from that observed at 1 AU, with a much more homogenous sheath. Indeed, the larger decrease in temperature in the sheath observed at Earth is due to the sheath being made of a ``new" sheath (shocked by the new shock) and an ``old" sheath (twice shocked). Having spacecraft closer to the Sun than 1 AU will provide evolutionary evidence of CMEs interaction. Here, having both observations at 0.33 and 1 AU, we can conclude that the trailing shock propagates inside the cloud and, consequently,  the single shock observed at 1 AU is the result of the merging of multiple shocks. Also the part of the first cloud just downstream of the trailing shock has already the same speed as that of the second cloud, whereas the part of the first cloud not yet shocked is slower. We can conclude from this observation that the trailing shock is primarily responsible for the acceleration of the first cloud. Consequently, the uniform speed at 1 AU is not due to momentum transfer between the clouds, but it is due to the acceleration of the cloud by the trailing shock and the presence of the second cloud which prevents the first one to decelerate after the shock's passage. This conclusion can only be reached by having in situ observations of the trailing shock propagating inside the first magnetic cloud, as well as observations after the interaction between the ejections took place.

\section{Coronagraphic Observations by STEREO}\label{STEREO}

The STEREO mission has been successfully launched in October 2006. In this section, we consider synthetic observations by the Heliospheric Imagers 1 and 2 ({\it HI1} and {\it HI2}) instruments onboard the two satellites. The {\it HI1} instruments have a field-of-view of $20^\circ$ with an angular offset of $13.65^\circ$ towards the Sun-Earth line and the {\it HI2} instruments have a field-of-view of $69^\circ$ with an angular offset of $53.35^\circ$ towards the Sun-Earth line.  Figure \ref{fig:HI1} shows different synthetic white-light images based on Simulation 2 corresponding to the satellites' configuration one year after the launch (two spacecraft separated by $45^\circ$).

The top panels of Figure \ref{fig:HI1} show synthetic images similar to what {\it HI1} would have observed. Here, at time 22 hours, the three leading edges can be observed and distinguished in both {\it HI1}'s fields-of-view at the same time. The 3-D structure of the CMEs is clearly visible. On the top right panel, the two leading edges have started interacting in HI1-B's field-of-view. This interaction is associated with an increase in the brightness, as the second shock propagating into the first dense sheath leads to a very large density increase. This is similar to what was observed in June 11, 1998 by LASCO C2 (see Figure \ref{fig:LASCO1998}).

The bottom panels of Figure \ref{fig:HI1} show synthetic images similar to what {\it HI2} would have observed at three different times. The top and bottom left panels show the ejections at the same time in {\it HI1} and {\it HI2} fields-of-view.
The brightness enhancement at the back of the first leading edge in the {\it HI2-B} is due to the second shock propagating into the leading edge of the first ejecta. Because {\it HI1} and {\it HI2} have different viewing angles, this feature is not yet observable in the {\it HI1} plots. The middle bottom panel shows the effect of the different propagation speeds in the different quadrant of the heliosphere, as {\it HI2-B} observes the large brightness increase associated with the interaction of the first two ejections, whereas {\it HI2-A} still observes two distinct ejecta. Six hours later (right panel), the interaction region between the two shocks is visible in {\it HI2-A} as well.

\section{Conclusion and Discussion}\label{conclusion}

We presented observational evidence of the interaction of multiple CMEs between the Sun and the Earth based on two 3-D MHD simulations. 
The main indicators of the interaction between multiple CMEs are the large temperature jump at the front shock and in the sheath and the high plasma $\beta$ in the region between the two clouds. The first feature indicates that the shock in front of the multiple-magnetic cloud event is the result of the merging of two shocks. The second feature indicates that reconnection happened between the two clouds. The short duration of the first cloud is also an indication that it has been compressed by an overtaking shock and that the natural tendency of the cloud to expand has been prevented by the presence of another cloud at its rear. 

Having spacecraft dedicated to the study of space weather in orbits close to the Sun will provide additional evidence of shocks propagating inside magnetic clouds and of shocks' merging. 
Furthermore, they could help validate and test numerical models. Also, we predict that the Heliospheric Imagers onboard STEREO will provide coronagraphic images of the interaction of leading edges of CMEs propagating towards Earth. This interaction is associated with a brightness increase and CME ``cannibalism", similar to what is observed by LASCO. Here we do not present synthetic LASCO observations, but the results found for the STEREO Heliospheric Imagers are directly applicable to LASCO. CME ``cannibalism" and shock-shock interaction in the field-of-view of LASCO C3 have also been numerically investigated by \citet{Schmidt:2004}. Also not presented here are synthetic in situ measurements at 1 AU but at different angles from the Sun-Earth line, as STEREO-A and B will provide. Because of the 3-D nature of CMEs and their shocks, STEREO A and B, {\it ACE} and {\it Wind} will be able to observe different phases of the interaction of multiple CMEs, which will also be very useful to validate and test 3-D numerical models. \\

\section*{Acknowledgments}
The research for this
manuscript was supported by Department of Defense MURI grant
F004449 and ITR grant 0325332 at the University of Michigan. We thank the referees for useful comments.
SOHO is a project of international cooperation between ESA and NASA. \\
W.~M. has been partially supported by the grants NASA-NNX06AC36G and LWS03-0130-0149 in subcontract from George Mason. I.~R. has been partially supported by the NSF grant ATM-0631790 (SHINE) and the Bulgarian NSF grant VU-01/06. N.~L. would like to thank NSF for travel grants to the ISROSES meeting.

\end{document}